\documentclass[nofootinbib,aps,superscriptaddress,preprint]{revtex4}

\usepackage{graphicx}

\def\d{{\rm d}}
\def\lqcd{\Lambda_{\rm QCD}}
\def\D0bar{\overline D{}^0}
\def\K0bar{\overline K{}^0}

\newcommand{\beq}{\begin{equation}}
\newcommand{\eeq}{\end{equation}}
\newcommand{\beqa}{\begin{eqnarray}}
\newcommand{\eeqa}{\end{eqnarray}}
\newcommand{\nn}{\nonumber}

\begin{document}

\preprint{\vbox {\hbox{SLAC-PUB-10342} \hbox{LBNL--54319}
  \hbox{WIS/05/04-Feb-DPP} \hbox{WSU--HEP--0401} \hbox{hep-ph/0402204}}}

\vspace*{2cm}

\title{\boldmath The $D^0-\D0bar$ mass difference from a dispersion relation}

\author{Adam F.\ Falk}
\affiliation{Department of Physics and Astronomy, 
	The Johns Hopkins University\\[-6pt]
	3400 North Charles Street, Baltimore, MD 21218}

\author{Yuval Grossman}
\affiliation{Department of Physics, Technion--Israel Institute of
Technology\\[-6pt]
	Technion City, 32000 Haifa, Israel}

\affiliation{Stanford Linear Accelerator Center\\[-6pt]
	Stanford University, Stanford, CA 94309}

\affiliation{Santa Cruz Institute for Particle Physics\\[-6pt]
	University of California, Santa Cruz, CA 95064}

\author{Zoltan Ligeti}
\affiliation{Ernest Orlando Lawrence Berkeley National Laboratory\\[-6pt]
	University of California, Berkeley, CA 94720}

\author{Yosef Nir}
\affiliation{Department of Particle Physics\\[-6pt]
	Weizmann Institute of Science, Rehovot 76100, Israel}

\author{Alexey A.\ Petrov\vspace{8pt}}
\affiliation{Department of Physics and Astronomy\\[-6pt]
	Wayne State University, Detroit, MI 48201\\[-6pt] $\phantom{}$ }

\begin{abstract}

We study the Standard Model prediction for the mass difference between the two
neutral $D$ meson mass eigenstates, $\Delta m$.  We derive a dispersion
relation based on heavy quark effective theory that relates $\Delta m$ to an
integral of the width difference of heavy mesons, $\Delta\Gamma$, over varying
values of the heavy meson mass.  Modeling the $m_D$-dependence of certain $D$
decay partial widths, we investigate the effects of $SU(3)$ breaking from phase
space on the mass difference.  We find that $\Delta m$ may be comparable in
magnitude to $\Delta\Gamma$ in the Standard Model.

\end{abstract}

\def\thepage{{}}
\maketitle

\def\thepage{\arabic{page}}

\section{Introduction}

The mixing and decay of $K$, $B$, and $D$ mesons are sensitive probes of
physics beyond the Standard Model.  Among the many processes that one might
study, flavor-changing neutral current $D$ decays and $D^0-\D0bar$ mixing
provide unique information, because in the Standard Model (SM) they occur via
loop diagrams involving intermediate down-type quarks.  In particular, because
of severe CKM and GIM suppressions, the mixing of $D$ mesons is expected to be
quite slow, and thus the $D$ system is one of the most intriguing probes of new
physics in low energy experiments~\cite{Burdman:2003rs}.

We begin by recalling the formalism for heavy meson mixing.  Using standard
notation, the expansion of the off-diagonal terms in the neutral $D$ mass
matrix to second order in perturbation theory is given by
\beq\label{M12}
\left (M - \frac{i}{2}\, \Gamma\right)_{12} =
  \frac{1}{2m_D}\, \langle D^0 | {\cal H}_w^{\Delta C=2} | \D0bar \rangle +
  \frac{1}{2m_D}\, \sum_n {\langle D^0 | {\cal H}_w^{\Delta C=1} | n 
  \rangle\, \langle n | {\cal H}_w^{\Delta C=1} | \D0bar \rangle 
  \over m_D-E_n+i\epsilon} \,.
\eeq
The first term represents the $\Delta C=2$ contributions that are local at the
scale $\mu \sim m_D$.  It contributes only to $M_{12}$, and is expected to be
very small unless it receives large enhancement from new physics.  The second
term in Eq.~(\ref{M12}) comes form double insertion of $\Delta C=1$ operators
in the SM Lagrangian and it contributes to both $M_{12}$ and $\Gamma_{12}$.  It
is dominated by the SM contributions even in the presence of new physics.  Two
physical parameters that characterize the mixing are
\beq
x = {\Delta m \over \Gamma}, \qquad 
y = {\Delta \Gamma \over 2\Gamma},
\eeq
where $\Delta m$ and $\Delta \Gamma$ are the mass and width differences of the
two neutral $D$ meson mass eigenstates and $\Gamma$ is their average width.
Because of the GIM mechanism the mixing amplitude is proportional to
differences of terms suppressed by $m_{d,s,b}^2/m_W^2$, and so $D^0-\D0bar$
mixing is very slow in the SM~\cite{Petrov:2003un}. The contribution of the $b$
quark is further suppressed by the small CKM elements $|V_{ub}
V_{cb}^*|^2/|V_{us} V_{cs}^*|^2 = {\cal O}(10^{-6})$, and can be neglected. 
Thus, the $D$ system essentially involves only the first two generations, and
therefore $CP$ violation is absent both in the mixing amplitude and in the
dominant tree-level decay amplitudes, and will be neglected hereafter.  Once
the contribution of $b$ quarks is neglected, the mixing vanishes in the flavor
$SU(3)$ limit, and it only arises at second order in $SU(3)$ breaking if
$SU(3)$ breaking can be treated analytically~\cite{Falk:2001hx}
\beq\label{theorem}
x\,,\, y \sim \sin^2\theta_C \times [SU(3) \mbox{ breaking}]^2\,,
\eeq
where $\theta_C$ is the Cabibbo angle.  Precise calculations of $x$ and $y$ in
the SM are not possible at present, because the charm mass is neither heavy
enough to justify inclusive calculations, nor is it light enough to allow a few
exclusive channels to give a reliable estimate.

According to Eq.~(\ref{theorem}), computing $x$ and $y$ in the SM requires a
calculation of $SU(3)$ violation in the decay rates.  There are many sources of
$SU(3)$ violation, most of them involving nonperturbative physics in an
essential way.  In Ref.~\cite{Falk:2001hx}, $SU(3)$ breaking arising from phase
space differences was studied; computing them in two-, three-, and four-body
$D$ decays, it was found that $y$ could naturally be at the level of one
percent. This result can be traced to the fact that the $SU(3)$ cancellation
between the contributions of members of the same multiplet can be badly broken
when decays to the heaviest members of a multiplet have small or vanishing
phase space.  This effect is manifestly not included in the OPE-based
calculations of $D^0-\D0bar$ mixing, which cannot address threshold effects.

The purpose of the present paper is to address the following question: if the
dominant $SU(3)$ breaking mechanism is indeed the one studied in
Ref.~\cite{Falk:2001hx}, and it gives rise to $y$ at the percent level, then
can $x$ naturally be comparably large?  This is particularly relevant because
the present experimental upper bounds on $x$ and $y$ are at the few times
$10^{-2}$ level~\cite{data1,data2} and are expected to significantly improve
(for a review of the experimental situation, see Ref.~\cite{Yabsley:2003rn}). 
To interpret the results from future measurements of $x$ and $y$, and possibly
establish the presence of new physics, we need to know the allowed range in the
SM.  In particular, since new physics can only contribute to $x$, an
experimental observation of $x \gg y$ would imply a large new physics
contribution to $D^0-\D0bar$ mixing. Although $y$ is determined by SM
processes, its value still affects the sensitivity to new
physics~\cite{Bergmann:2000id}.

In this paper we study the SM predictions for $x/y$ due to $SU(3)$ breaking
from final state phase space differences.  In Sec.~\ref{sec:form} we derive a
dispersion relation using Heavy Quark Effective Theory (HQET) that relates
$\Delta m$ to $\Delta\Gamma$.  To compute $\Delta m$, we need a calculation of
$\Delta\Gamma$ for varying heavy meson mass, so we review its calculation from
Ref.~\cite{Falk:2001hx} in Sec.~\ref{sec:y}.  In Sec.~\ref{sec:calc}, we
calculate $\Delta m$ and present numerical results.  We find that despite the
fact that $SU(3)$ breaking in phase space affects $x$ in a different way than
it affects $y$, the final estimates of $x$ and $y$ are comparable.   We present
our conclusions in Sec.~\ref{sec:concl} and discuss the implications of our
findings for experimental searches for new physics in $D^0-\D0bar$ mixing.

\section{Derivation of the dispersion relation}
\label{sec:form}

We start by reviewing the relevant formalism for $D^0-\D0bar$ mixing. 
Equation~(\ref{M12}) implies that the mass eigenstates are linear combinations
of the weak interaction eigenstates, $| D_{1,2} \rangle = p\, | D^0 \rangle \pm
q\, | \D0bar \rangle$. Since we neglect the effects of intermediate states
containing a $b$ quark, $| D_{1,2} \rangle$ are also $CP$ eigenstates, $CP |
D_{\pm} \rangle = \pm | D_{\pm} \rangle$. Their mass and width differences are
\beq
\Delta m \equiv m_{D_+} - m_{D_-} = 2 M_{12} \,, \qquad
  \Delta \Gamma \equiv \Gamma_{D_+} - \Gamma_{D_-} = 2 \Gamma_{12} \,.
\eeq
Neglecting the small contribution from the local $\Delta C=2$ operators,
Eq.~(\ref{M12}) gives
\beqa\label{xy}
\Delta m &=& {1\over 2m_D}\, {\rm P} \sum_n 
  {\langle D^0 | {\cal H}_w | n \rangle \langle n | {\cal H}_w | \D0bar \rangle
  + \langle \D0bar | {\cal H}_w | n \rangle
  \langle n | {\cal H}_w| D^0 \rangle \over m_D - E_n}\,, \nn\\
\Delta \Gamma &=& {1\over 2m_D}\, \sum_n \Big[ 
  \langle D^0 | {\cal H}_w | n \rangle \langle n | {\cal H}_w| \D0bar \rangle 
  + \langle \D0bar | {\cal H}_w | n \rangle
  \langle n | {\cal H}_w | D^0 \rangle \Big] (2\pi)\delta(m_D-E_n) \,,
\eeqa
where P denotes the principal value prescription, the sum is over all
intermediate states, $n$, and it implicitly includes $(2\pi)^3 \delta^3(\vec
p_D - \vec p_n)$.

\begin{figure}[t]
\centerline{\includegraphics[width=.5\textwidth]{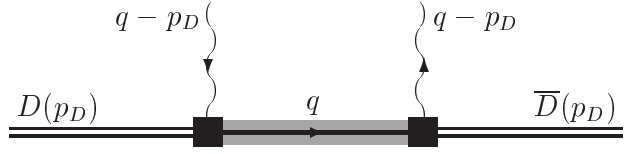}}
\caption{The correlator in Eq.~(\ref{basdef}).  The black boxes denote the weak
Hamiltonian, the wavy lines show external momenta inserted, and the gray area
represents hadronic intermediate states.}
\label{fig:corr}
\end{figure}

To derive a dispersion relation between $\Delta m$ and $\Delta\Gamma$, consider
the following correlator
\beq\label{basdef}
\Sigma_{p_D}(q) = i \int\! \d^4z\, \langle \overline D(p_D)|\,
  T\left[ {\cal H}_w(z)\, {\cal H}_w(0)\right] |D(p_D)\rangle\,
  e^{i(q-p_D)\cdot z}\,.
\eeq
Here $p_D$ is a label given by the momentum of the on-shell $D$ meson state
(satisfying $p_D^2=m_D^2$) and $q-p_D$ is an auxiliary four-vector that inserts
external momentum to the weak interaction (see Fig.~\ref{fig:corr}). There is
no simple physical interpretation of $\Sigma$ except at $q = p_D$, where
$\Sigma_{p_D}(p_D)$ is related to physical properties of $D$ mesons.  Inserting
a complete set of states in Eq.~(\ref{basdef}) and comparing with
Eq.~(\ref{xy}), we find
\beq\label{phyint-pre}
- {1\over 2m_D}\, \Sigma_{p_D}(p_D) = \bigg( 
  \Delta m - {i\over 2}\,\Delta\Gamma \bigg) \,.
\eeq
The correlator $\Sigma_{p_D}(q)$ is an analytic function of $q$ (but not of
$p_D$) with a cut in the complex $q^0$ plane for $q^0 > \sqrt{ |\vec q\,|^2 +
4m_\pi^2}$ for a fixed $\vec q$.

To write the dispersion relation in terms of physical quantities, i.e., to give
$\Sigma_{p_D}(q)$ for $q \neq p_D$ a physical interpretation, we need to
eliminate the heavy quark mass dependence from Eq.~(\ref{basdef}).\footnote{The
method of using HQET to derive a dispersion relation in the heavy quark mass
was developed first in Ref.~\cite{Grinstein:2001nu}, where it was used to study
the inclusive nonleptonic heavy meson decay rate.}  The momentum of a heavy
meson $H$ containing a heavy quark $Q$ can be written as $p_H^\mu = m_H v^\mu$,
with $v^2=1$.  We can decompose $Q$ as
\beq\label{hqetf}
Q(z) = e^{-im_Q v\cdot z} h_v^{(Q)}(z) 
  + e^{+im_Q v\cdot z} \widetilde h_v^{(Q)}(z) + \ldots \,,
\eeq
where the HQET fields $h_v^{(Q)}$ and $\widetilde h_v^{(Q)}$ respectively
annihilate a heavy $Q$ quark and create a heavy $\bar Q$ antiquark with
four-velocity $v$.  Here and in the rest of this section the ellipses denote
terms suppressed by a relative factor of $\lqcd/m_c$.  The $\Delta C = 1$ weak
Hamiltonian contributing to neutral $D$ meson mixing is
\beq\label{hqeth}
{\cal H}_w = {4G_F \over\sqrt2}\, V_{cq_1} V_{uq_2}^*\, \sum_{i} C_i\, O_i
  = \hat {\cal H}_w \Big[ e^{-im_c v\cdot z}\, h_v^{(c)}
  + e^{im_c v\cdot z}\, \widetilde h_v^{(c)} \Big] + \ldots \,,
\eeq
where $q_{1,2}=d$ or $s$, and the four-quark operators, suppressing their Dirac
structures, are of the form
\beq
O_i \sim \bar q_1 q_2\bar u c =
  e^{-im_c v\cdot z}\, \bar q_1 q_2\bar u h_v^{(c)}
  + e^{im_c v\cdot z}\, \bar q_1 q_2\bar u\widetilde h_v^{(c)} + \ldots \,.
\eeq
In Eq.~(\ref{hqeth}) $\hat {\cal H}_w$ contains the light quark fields, the
Wilson coefficients, and summation over operators.  We also replace the QCD
states $|D\rangle$ by HQET states $|H(v)\rangle$,
\beq
|D(p=m_Dv)\rangle = \sqrt{m_D}\, |H(v)\rangle + \ldots \,.
\eeq
The new states have a normalization that is independent of the heavy quark
mass~\cite{Manohar:dt}.  Then Eq.~(\ref{basdef}) yields
\beqa\label{hqetsig}
\Sigma_{p_D}(q) &=& i\, m_D \int\! \d^4z\, \langle \overline H(v)|\,
  T\, \Big\{\Big[ e^{-im_c v\cdot z}\, \hat {\cal H}_w h_v^{(c)}(z)+
  e^{im_c v\cdot z}\, \hat {\cal H}_w \widetilde h_v^{(c)}(z)\Big] \nn\\
&&{}\qquad\quad \times \Big[ \hat {\cal H}_w h_v^{(c)}(0) +
  \hat {\cal H}_w \widetilde h_v^{(c)}(0) \Big]\Big\}\,
  e^{i(q-p_D)\cdot z}\, |H(v)\rangle + \ldots\,.
\eeqa
The only nonzero contributions to this correlator involve a single $h$ and
$\widetilde h$ field each,
\beqa\label{hqesig}
\Sigma_{p_D}(q) &=& i\,m_D \int\! \d^4z\, \langle\overline H(v)|\,
  \Big\{ e^{-im_c v\cdot z}\, T \Big[\hat {\cal H}_w h_v^{(c)}(z),
  \hat {\cal H}_w \widetilde h_v^{(c)}(0)\Big] \nn\\
&&{}\qquad + e^{im_c v\cdot z}\,
  T \Big[\hat {\cal H}_w \widetilde h_v^{(c)}(z),
  \hat {\cal H}_w h_v^{(c)}(0)\Big]\Big\}\, e^{i(q-p_D)\cdot z}\,|H(v)\rangle
  + \ldots\,.
\eeqa

The two terms in Eq.~(\ref{hqesig}) behave differently in the HQET limit $m_c
\to \infty$ with $q$ fixed.  The term proportional to  $\exp[i(q-p_D-m_c
v)\cdot z]$ oscillates infinitely rapidly and is integrated out at the heavy
scale.  It should be removed  from the effective theory and replaced by a local
${\cal H}_w^{\Delta  C=2}$ contribution that can be included as a matrix
element of $\Delta C=2$ operators.  Such contributions are estimated to give
rise to $x$ and $y$ at or below the $10^{-3}$
level~\cite{Georgi:1992as,Ohl:1992sr,Bigi:2000wn},\footnote{In the  OPE-based
calculations, because $m_c/\lqcd$ is not very large and subleading  terms in
the $\lqcd/m_c$ expansion are enhanced by $\Lambda_{\chi
SB}/m_s$~\cite{Georgi:1992as}, such terms dominate the short distance
contribution~\cite{Georgi:1992as,Ohl:1992sr,Bigi:2000wn}.} and since we  are
interested in the question whether $x$ could be near the percent level,  we can
neglect them.

By contrast, the term proportional to $\exp[i(q-p_D+m_c v)\cdot z]$ becomes
independent of $m_c$ as $m_c\to\infty$.  Recalling that $p_D=m_Dv$, we have
\beq\label{hqsig}
\Sigma_{p_D}(q) = i\, m_D \int\! \d^4z\, \langle\overline H(v)|\,
  T \Big[\hat {\cal H}_w \widetilde h_v^{(c)}(z), 
  \hat {\cal H}_w h_v^{(c)}(0)\Big]\,
  e^{i(q-\bar\Lambda v)\cdot z}\, |H(v)\rangle + \ldots\,,
\eeq
where $\bar\Lambda = m_D - m_c + {\cal O}(\lqcd^2/m_c)$.  It is convenient to
define
\beq\label{barsigsig}
\overline\Sigma_v(q) = i \int\! \d^4z\, \langle\overline H(v)|\,
  T \Big[ \hat {\cal H}_w \widetilde h_v^{(Q)}(z), 
  \hat {\cal H}_w h_v^{(Q)}(0) \Big]\, 
  e^{i(q-\bar\Lambda v)\cdot z}\, |H(v)\rangle \,,
\eeq
which is manifestly independent of the heavy quark mass.  It follows that 
\beq\label{sigrel}
\Sigma_{p_D}(q) = m_D\, \overline\Sigma_v(q) + \ldots \,,
\eeq
and Eq.~(\ref{phyint-pre}) becomes to leading order in $\lqcd/m_c$
\beq\label{phyint}
\overline\Sigma_v(q) = -2\, \Delta m(E) + i\, \Delta\Gamma(E)\,,
\eeq
where $E \equiv \sqrt{q^2}$, and $\Delta m(E)$ and $\Delta\Gamma(E)$ can be
interpreted as the mass and the width differences of neutral heavy mesons with
mass $E$ in HQET.  Equation~(\ref{phyint}) shows that $\overline\Sigma_v(q)$
only depends on $q^2$.  Choosing a frame in which $\vec q = 0$, we can use the
analyticity of $\overline\Sigma_v(q)$ to write a dispersion relation,
\beq\label{disbar}
\overline\Sigma_v(m_D,\vec 0) = \frac1\pi\, \int_{2m_\pi}^\infty \d E\,
  \frac{{\rm Im}\, \overline\Sigma_v(E,\vec 0)}{E-m_D+i\epsilon}\,.
\eeq
Using Eq.~(\ref{phyint}), we obtain
\beq\label{delmd}
\Delta m = -\frac{1}{2\pi}\, {\rm P}\! \int_{2m_\pi}^\infty \d E
  \left[ \frac{\Delta\Gamma(E)}{E-m_D} + {\cal O}\bigg( {\lqcd\over E} \bigg) 
  \right] .
\eeq

Eq.~(\ref{delmd}) is the main result of this section.  It expresses $\Delta
m_D$ in terms of a weighted integral of the width difference of heavy mesons,
$\Delta\Gamma(E)$, over varying heavy meson masses, $E$.  The heavy quark limit
was essential in deriving this relation, since $\overline\Sigma_v(q)$ has a
physical interpretation for arbitrary $q$, while for $q \neq p_D$,
$\Sigma_{p_D}(q)$  does not.  The ${\cal O}(\lqcd/E)$ error in the integrand is
a consequence of our reliance on this limit, and the resulting correction is
${\cal O}(1)$ in the small $E$ region.  Dispersion relations for $\Delta m_D$
were considered previously in Ref.~\cite{Donoghue:hh}, where ${\rm Im}\,
\Sigma(s)$ (with a different definition of $\Sigma$) was modeled, but it does
not have a physical interpretation for $s \neq m_D^2$.

To calculate $x/y$ using the dispersion relation, we need to know
$\Delta\Gamma$ as a function of the heavy meson mass.  Examining
Eq.~(\ref{delmd}),  we expect that values of $E$ close to $m_D$ give the
largest contribution to $x$.  In the next section we recall the calculation of
$\Delta\Gamma(E)$ performed in Ref.~\cite{Falk:2001hx}.  If $\Delta\Gamma(E)$
is a decreasing function of $E$ at least as a positive power, $1/E^a$ with
$a>0$, then the dispersion relation does not require subtraction in order to
converge.  In the model we consider, $\Delta\Gamma(E)$ actually falls off as
$\sim 1/E^2$, and we will argue that some kind of decreasing behavior is likely
to hold model independently.

\section{Calculation of the lifetime difference}
\label{sec:y}

The computation of $x$ using Eq.~(\ref{delmd}), requires us to know $\Delta
\Gamma$ for a heavy meson of varying mass.  The calculation of $\Delta\Gamma$
cannot at present be done from first principles.  In Ref.~\cite{Falk:2001hx}
$\Delta\Gamma$ was computed using a simple model in which $SU(3)$ breaking was
taken into account in calculable phase space differences, but neglected in the
incalculable hadronic matrix elements.  This approach was motivated by the fact
that phase space differences alone can explain the experimental data in several
cases; for example the ratio $\Gamma(D_2^*\to D\pi) / \Gamma(D_2^*\to
D^*\pi)$~\cite{Isgur:wq}, the large $SU(3)$ breaking in $\Gamma(D\to
K^*\ell\bar\nu) / \Gamma(D\to \rho\ell\bar\nu)$~\cite{Ligeti:1997aq}, and the
lifetime ratio $\tau_{D_s} / \tau_{D^0}$~\cite{Nussinov:2001zc}.  It certainly
cannot explain all $SU(3)$ violation, for example, $\Gamma(D\to\pi\pi) /
\Gamma(D\to KK)$.  The generic conclusion of Ref.~\cite{Falk:2001hx} was that
if multi-body final states close to the $D$ threshold have significant
branching ratios, then they can give rise to sizable contributions to
$\Delta\Gamma$ that are absent in the OPE-based calculations.  Our purpose in
the next section will be to see whether the same mechanism can also give rise
to $x$ at or near the percent level.  Here we review the analysis of
Ref.~\cite{Falk:2001hx}.

We denote a set of final states $F$ belonging to a certain representation $R$
of $SU(3)$ by $F_R$.  For example, for two pseudoscalar mesons in the octet,
the possible representations for $F=PP$ are $R=8$ and $27$.  In
Ref.~\cite{Falk:2001hx} it was shown that $y_{F_R}$, the value which $y$ would
take if elements of $F_R$ were the only channels open for $D^0$ decay, can be
expressed as
\beq\label{yfr}
y_{F_R} = {\sum_{n\in F_R} \langle\D0bar| {\cal H}_w|n\rangle
  \langle n| {\cal H}_w |D^0\rangle \over
  \sum_{n\in F_R} \langle D^0| {\cal H}_w |n\rangle
  \langle n|{\cal H}_w |D^0\rangle} 
= {\sum_{n\in F_R} \langle\D0bar| {\cal H}_w|n\rangle
  \langle n| {\cal H}_w |D^0\rangle \over \sum_{n\in F_R} \Gamma(D^0\to n)} \,.
\eeq
The derivation of this relation assumes the absence of $CP$ violation, so that
$\langle\D0bar|{\cal H}_w|n\rangle$ is related to $\langle D^0|{\cal H}_w|\bar
n\rangle$, and uses the fact that both $|n\rangle$ and $|\bar n\rangle$ belong
to the same $SU(3)$ multiplet.  When the $SU(3)$ breaking in the matrix
elements is neglected, Eq.~(\ref{yfr}) gives a calculable contribution to
$y_{F_R}$ without any hadronic parameters.  The numerator contains a
combination of Clebsch-Gordan and CKM coefficients that ensures that $y_{F_R}$
is proportional to $m_s^2\, \sin^2\theta_C$ when the sum over all members of
any given multiplet $F_R$ is performed, as required by Eq.~(\ref{theorem}).

As an example, the contribution of the multiplet containing two pseudoscalar
mesons in an $SU(3)$ octet is given by
\beqa\label{yPP8}
y_{(PP)_8} &=& \sin^2\theta_C \bigg[ \frac12\, \Phi(\eta,\eta)
  + \frac12\,  \Phi(\pi^0,\pi^0) + \Phi(\pi^+,\pi^-) + \Phi(K^+,K^-) \nn\\*
&&{} \qquad\quad + \frac13\, \Phi(\eta,\pi^0) - \frac13\, \Phi(\eta,K^0) 
  - 2\Phi(K^+,\pi^-) - \Phi(K^0,\pi^0) \bigg] \nn\\*
&&{} \qquad \times \bigg[ \frac16\, \Phi(\eta,\K0bar) + \Phi(K^-,\pi^+)
  + \frac12\, \Phi(\K0bar,\pi^0)\bigg]^{-1} + {\cal O}(\sin^4\theta_C) \,,
\eeqa
where $\Phi(n)$ is the phase space factor for $D\to n$ decay.  Then $y$ can be
computed as the sum of the $y_{F_R}$'s weighted with the $D^0$ decay rate to
each representation,
\beq\label{ycombine}
y = {1\over\Gamma} \sum_{F_R}\, y_{F_R}
  \bigg[ \sum_{n\in F_R} \Gamma(D^0\to n) \bigg]\,.
\eeq
The $y_{F_R}$ were computed for all $PP$, $PV$, and $VV$ representations, and
for the fully symmetric $3P$ and $4P$ final states~\cite{Falk:2001hx}.  The
contribution of poles corresponding to nearby $K$ resonances was shown to be
small~\cite{Golowich:1998pz,Falk:2001hx}.  Assuming that the values of
$y_{(4P)_R}$ for $R=8,\ 27,\ 27'$ are typical for all $R$, it was found that
the $4P$ final states give a contribution to $\Delta\Gamma$ at the percent
level.  The result is large because many of the decays in question are close to
or above threshold, so the $SU(3)$ cancellation in these multiplets is largely
ineffective, yielding $y_{(4P)_R} = {\cal O}(0.1)$~\cite{Falk:2001hx}. 
Moreover, the $D^0$ branching ratio to four pseudoscalars is approximately
10\%.

We shall now use this model of $SU(3)$ breaking, together with some assumptions
about the energy dependences of the relevant decay rates, to compute $x/y$.

\section{Calculation of the mass difference}
\label{sec:calc}

The crucial difference between the calculation of $x$ and $y$ is that once we
assume that the only source of $SU(3)$ breaking is from the final state phase
space differences, the hadronic matrix elements cancel in $y$, but not in $x$.
As determined by Eq.~(\ref{delmd}), $x$ depends on $\Delta\Gamma(E)$, and so
the $E$-dependence of the hadronic matrix elements does affect $x$.  Using
Eq.~(\ref{delmd}), we find for $x/y$,
\beq\label{xovery}
r_{F_R} \equiv{x_{F_R} \over y_{F_R}} 
= -\frac{1}{\pi}\, {\rm P}\! \int_{2m_\pi}^\infty 
  \frac{\d E}{E-m_D}\, \frac{y_{F_R}(E)}{y_{F_R}(m_D)}\,
  {\Gamma_{F_R}(E) \over \Gamma_{F_R}(m_D)} \,.
\eeq
We will quote our results in terms of $r_{F_R}$.  To proceed further we need to
understand or make some assumptions about the $E$-dependence of the decay rate
to the final state $F$, $\Gamma_F(E)$.  We define the dimensionless function
\beq\label{gdef}
g_F(E) \propto {\Gamma_F(E) \over \Gamma_F(m_D)}\,,
\eeq
and we will study the $E$-dependence of this quantity.  Note that the constant
of proportionality in Eq.~(\ref{gdef}) cancels in the ratio $r_{F_R}$. 
Moreover, $g_F$ is expected to depend only on the final state $F$, and not on
the $SU(3)$ representation $R$.

One can reconstruct $x$ from $x_{F_R}$ using a relation analogous to
Eq.~(\ref{ycombine}).  Below we calculate $r_{F_R}$ for several final states
and then estimate the total $x$.  First we will study $F = PP$, because it is
a simple case that is interesting to understand in detail.  Then we will turn
to $F = 4P$, because it is the final state that can give $y \sim 1\%$.

\subsection{\boldmath Two-body $D\to PP$ decays}

For decays to two pseudoscalar mesons, it is possible to develop a reasonable
model of $g_{PP}(E)$.  When $m_H \gg\lqcd$, we may approximate the $H \to
\pi\pi$ amplitude with its factorized form.  Here $A(H \to \pi\pi) \sim G_F
V_{\rm CKM}\, m_H^2\, f_\pi F_{H \to \pi}$, where $f_\pi$ is the pion decay
constant and $F_{H\to\pi}$ is the $H \to \pi$ form factor at $q^2=m_\pi^2$.  It
has been shown that, as $m_H\to\infty$, $F_{H\to\pi} \propto (\Lambda /
m_H)^{3/2+X}$~\cite{Chernyak:ag}, where $X$ arises from summing Sudakov
logarithms of the form $\exp [C \alpha_s(m_H) \ln^2(m_H/\Lambda)] \sim
(\Lambda/m_H)^X$ with $X=-2\pi C/\beta_0$.  Since $\Gamma \propto |A|^2/m_H$,
we conclude that
\beq\label{high-s}
g_{PP}(E \gg \lqcd) \propto E^{-2X} .
\eeq
The existing calculations suggest that $|X| \ll 1$~\cite{Akhoury:uw}, so we set
$X=0$ hereafter.

In the $E \to 0$ limit our calculation is necessarily unreliable, as the
derivation of Eq.~(\ref{delmd}) relied on HQET.  Nevertheless, as a model we
will take the behavior of the $K \to \pi\pi$ amplitude in chiral perturbation
theory.  At leading order, this transition is mediated by an operator of the
form ${\rm Tr}(\partial_\mu\Sigma^\dagger\, {\cal O}\, \partial^\mu\Sigma)$,
where $\Sigma = \exp[2i{\cal M}/f]$ and ${\cal M}$ is the meson octet.  Since
this term has two derivatives, it implies that the decay amplitude is
proportional to $m_K^2$.  Since this is the only dependence on $m_K$ in the
amplitude, the $E$-dependence of the rate is
\beq
g_{PP}(E \to 0) \propto E^3 .
\eeq

Based on these considerations, we employ the following simple model for
$g_{PP}(E)$
\beq\label{gppdef}
g_{PP}(E) = \cases{ E^3/(m_1^2 m_2)  &  for $E < m_1$\,, \cr
  E/m_2  &  for $m_1 < E < m_2$\,, \cr
  1  &  for $E > m_2$\,, \cr}
\eeq
where $m_{1,2}$ are free parameters. The overall normalization cancels in the
results.   This model allows for a ``chiral" region, $E < m_1$, an
``intermediate" region, $m_1 < E < m_2$, and a ``high energy" region, $E >
m_2$.  In our calculations $m_1$ is allowed to vary in the range
$0.2-1.0\,$GeV, and $m_2$ in the range $1.5-10\,$GeV.  As we emphasized above,
our derivation relies on HQET, so any strong dependence on scales below $\sim
1\,$GeV would signal an irreducible lack of reliability.

\begin{figure}[t]
\centerline{\includegraphics[width=.5\textwidth]{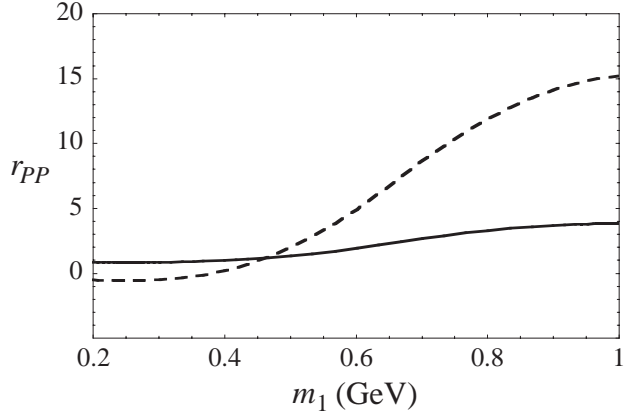}}
\caption{Predictions for $r_{(PP)_8}$ (solid curve) and $r_{(PP)_{27}}$ (dashed
curve) as functions of $m_1$.}
\label{fig:pp}
\end{figure}

In Fig.~\ref{fig:pp} we plot $r_{(PP)_8}$ (solid) and $r_{(PP)_{27}}$ (dashed)
as a function of $m_1$, for $m_2 = 2\,$GeV.  In this case all members of the
final state representations are kinematically allowed and have large phase
space, so we find that the result is dominated by cancellations below the scale
$m_D$.  Therefore $r_{PP}$ is sensitive to the shape of $g_{PP}(E)$ at low
energies, i.e., the value of $m_1$, but changing $m_2$ to 3 or 4\,GeV has
little effect on $r_{PP}$.  Because of the strong dependence on $m_1$, we
should not trust this result.  However, since $y$ for these representations is
very small, $y_{(PP)_8} = -0.018\%$ and $y_{(PP)_{27}} =
-0.0034\%$~\cite{Falk:2001hx}, these final states do not give sizable
contributions to $x$ in any case.

\begin{figure}[t]
\centerline{\includegraphics[width=.5\textwidth]{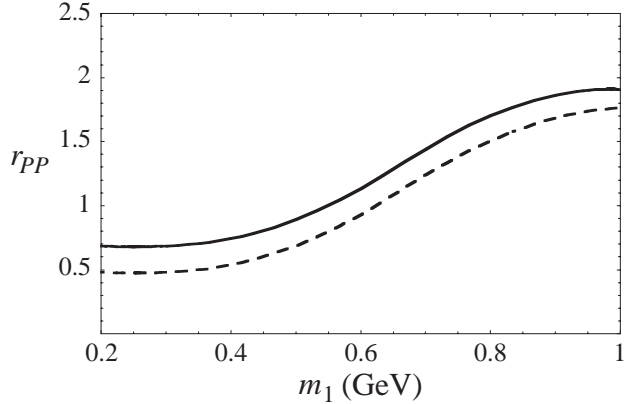}}
\caption{Predictions for $r_{(PP)_8}$ (solid curve) and $r_{(PP)_{27}}$ (dashed
curve) as functions of $m_1$, imposing the GMO relation on the $\pi$, $K$, and
$\eta$ masses.}
\label{fig:ppgmo}
\end{figure}

When we consider decays to the lightest pseudoscalar octet, the dependence of
these pseudo-Goldstone boson masses on $m_s$ is given by (for $m_{u,d} = 0$) 
\beq\label{gmorel}
m_\pi^2 = 0\,, \qquad m_K^2 = \mu m_s\,, \qquad m_\eta^2 = \frac43\, \mu m_s\,,
\eeq
where $\mu$ is a hadronic scale.  We can then expand $\Delta\Gamma(E)$ for
large $E$ as
\beq
\Delta\Gamma_{PP}(E) = \bigg[\Gamma_{PP}(E)\Big|_{m_s\to 0}\bigg] \times 
  \left(c_0 + \frac{c_1}{E^2} + \frac{c_2}{E^4} + \ldots \right).
\eeq
Because $SU(3)$ breaking in our approach comes from phase space differences,
the coefficients $c_i$ depend quadratically on the masses of the final state
particles.  Since in Eq.~(\ref{gmorel}) $m_s$ is always accompanied by $\mu$
and $\Delta\Gamma$ must be suppressed by $m_s^2$, we conclude that $c_0 = c_1 =
0$.  The coefficient $c_2$ can be proportional to $\mu^2 m_s^2$ and is the
leading nonvanishing term, implying a $1/E^4$ suppression of
$\Delta\Gamma_{PP}(E)$ compared to $\Gamma_{PP}(E)$.  However, the actual
$\pi$, $K$, and $\eta$ masses do not exactly satisfy Eq.~(\ref{gmorel}) in the
$m_{u,d} = 0$ limit, nor the Gell-Mann-Okubo (GMO) relation, $3 m_\eta^2 =
4m_K^2-m_\pi^2$.  Violating the GMO relation is equivalent to adding a small
term to $m_K^2$ or to $m_\eta^2$ of the form $\varepsilon\, m_s^2$.  This
changes the asymptotic behavior of $\Delta\Gamma(E)$, because now we can have
$c_1 \sim \varepsilon\,m_s^2$.  Since the $D \to PP$ decay is far from
threshold, the $SU(3)$ cancellation in this channel is very sensitive to the
pseudoscalar meson masses.  This can be verified analytically by expanding
Eq.~(\ref{yPP8}).  As shown in Fig.~\ref{fig:ppgmo} (again for $m_2 = 2\,$GeV),
imposing the GMO relation on the $\pi$, $K$, and $\eta$ masses decreases
$r_{PP}$ significantly, in such a manner that $y_{PP}$ increases by roughly the
same factor, while $|x_{PP}|$ is approximately stable at the $(5-8) \times
10^{-4}$ level.  As discussed in Ref.~\cite{Falk:2001hx}, our results have
little sensitivity to including or neglecting $\pi-\eta-\eta'$ mixing.

By contrast, for final states including vector mesons or heavier pseudoscalar
representations, the masses of the mesons depend linearly on $m_s$.  Thus, for
these final states, $\Delta\Gamma_F(E) / \Gamma_F(E)$ is simply proportional to
$m_s^2/E^2$ for large $E$, and there is no strong dependence on the precise
values of the hadron masses.  This is the minimal suppression of
$\Delta\Gamma_F(E) / \Gamma_F(E)$ consistent with group theory, i.e.,
Eq.~(\ref{theorem}), and our phase space model for $SU(3)$ violation indeed
gives such an effect.  These results imply that the dispersion relation in
Eq.~(\ref{delmd}) converges for any final state $F$, for which $\Gamma_F(E)$
does not increase as $E^2$ or faster.  This is very likely to be true for all
final states (recall that $\Gamma_{PP}(E) \sim \mbox{constant}$ for large
$E$).

\subsection{\boldmath Four-body $D\to 4P$ decays}

Now we turn to the $4P$ final state in the fully symmetric 8, 27, and $27'$
representations of $SU(3)$.  We know even less about $g_{4P}(E)$ than about
$g_{PP}(E)$, so we use two models to attempt to bracket roughly the
uncertainties,
\beq\label{g4pdef}
g_{4P}(E) = g_{PP}(E) \qquad \mbox{and} \qquad
g'_{4P}(E) = \cases{ E/m_1  &  for $E < m_1$\,, \cr
  1  &  for $m_1 < E < m_2$\,, \cr
  m_2/E  &  for $E > m_2$\,. \cr}
\eeq
The choice of $g'_{4P}(E)$ allows for the possibility that $\Gamma(H\to 4P)$
may start to fall for large $m_H$ instead of remain constant.  This alternative
is motivated by the argument that because the quasi-two-body picture holds only
in a small part of phase space, in most of the phase space the opening of many
decay channels will reduce the rate.

\begin{figure}[t]
\centerline{\includegraphics[width=.49\textwidth]{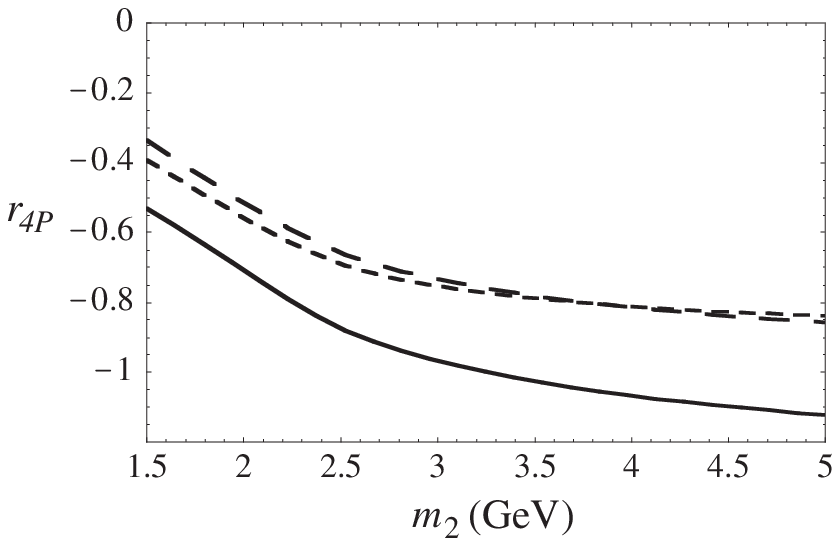}\hfill
\includegraphics[width=.49\textwidth]{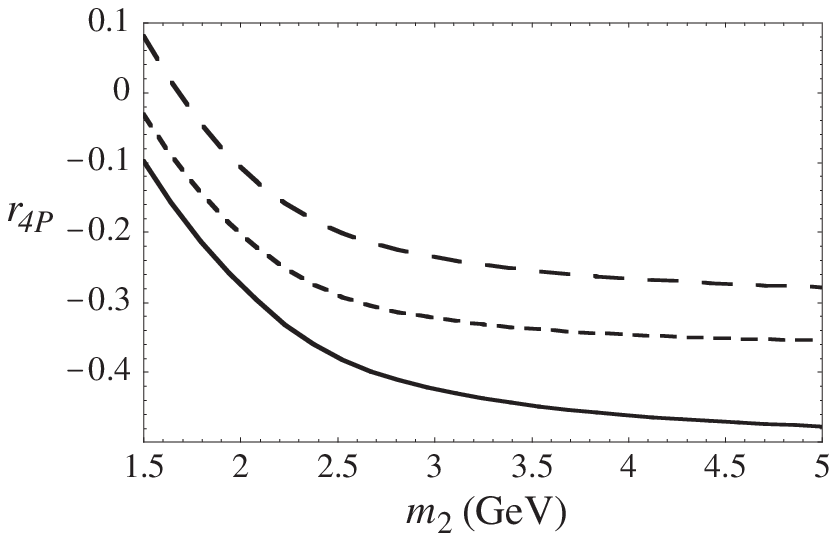}}
\caption{Predictions for $r_{(4P)_8}$ (solid curve), $r_{(4P)_{27}}$ (long
dashed curve), and $r_{(4P)_{27'}}$ (short dashed curve), as functions of $m_2$
for the models $g_{4P}(E)$ (left figure) and $g'_{4P}(E)$ (right figure) in
Eq.~(\ref{g4pdef}).}
\label{fig:pppp}
\end{figure}

The left plot in Fig.~\ref{fig:pppp} shows $r_{(4P)_8}$ (solid curve),
$r_{(4P)_{27}}$ (long dashed curve), and $r_{(4P)_{27'}}$ (short dashed curve),
as functions of $m_2$, using $g_{4P}(E)$ with $m_1 = 0.8\,$GeV.  For $m_1 < 1\,
{\rm GeV}$ there is no dependence on $m_1$.  The dependence of the curves on
$m_2$ is negligible for $m_2 \gtrsim 3\,$GeV.  If we use $g'_{4P}(E)$ instead,
shown in the right plot in Fig.~\ref{fig:pppp}, then $r_{(4P)_R}$ changes
roughly by a factor of two.  We have explored other forms of $g_{4P}(E)$ as
well, and we find that these two cases cover a reasonable range of predictions.

In contrast to $D\to PP$ decays, for the $4P$ final state there is no strong
dependence on the $\pi$, $K$ and $\eta$ masses.  Because the decay is close to
threshold, the dispersion integral is dominated by $E$ near $m_D$, where some
of the $4P$ final states are kinematically forbidden, and so the sensitivity to
the pseudoscalar meson masses is reduced.  Imposing the GMO relation makes only
a small difference; for example, for the $(4P)_8$ representation the value
$r_{(4P)_8} = -0.98$ obtained with the $g_{4P}(E)$ model, $m_1 = 0.8\,$GeV,
$m_2 = 3\,$GeV, and the physical meson masses (corresponding to the solid curve
in the left plot in Fig.~\ref{fig:pppp}), would change to $r_{(4P)_8} = -0.87$
if the GMO relation were imposed.

\section{Discussion and Conclusions}
\label{sec:concl}

It is likely that the dominant contributions to the mass and width differences
in the $D$ system have a long distance origin in the SM.  Therefore, naively
one would expect $x$ and $y$ to be of the same order of magnitude.  We have
derived a new dispersion relation (\ref{delmd}) and used it to study this
question.  Our dispersion relation has the useful property that it relates the
mass difference in the heavy neutral meson system at fixed heavy meson mass to
the physical width difference of heavy mesons with varying mass.  

The advantage of using a dispersion relation that relates $x$ to $y$ is that we
can use existing models for $y$ to calculate $x$.  Our dispersion relation is
likely to converge without any subtraction, because the $SU(3)$ breaking
required to yield nonzero mixing introduces an $m_s^2/E^2$ suppression in
$y(E)$.  We have used a model in which $SU(3)$ breaking arises from phase space
differences, which may give a reasonable approximation to $y(E)$ only when $E$
is not very large.  Since the derivation of the dispersion relation employed
the heavy quark limit, it is essential not to interpret our analysis as a
precise calculation for $x$.  Instead, we used this model only to get a rough
and qualitative prediction about the likely relation of $x$ to $y$.

To make numerical predictions we needed the heavy mass dependence of heavy
meson partial widths to certain final states, which introduces some additional
model dependence in our results.  (For decays to two pseudoscalars, there are
limits in which one can draw firmer conclusions about the mass dependence,
which we have incorporated into the model.)  We calculated the ratio $x/y$ for
$PP$ and $4P$ final states.  Our conclusion is that it is indeed likely that in
the Standard Model, $x$ is not much smaller than $y$ in the $D$ system.  In our
numerical study, we found that for the $4P$ final state, $x/y$ varies roughly
between $-0.1$ and $-1$.  We conclude that if $y$ is in the ballpark of $+1\%$
as expected if the $4P$ final states dominate $y$~\cite{Falk:2001hx}, then we
should expect $|x|$ between $10^{-3}$ and $10^{-2}$, and that $x$ and $y$ are
of opposite sign.  This estimate has a large uncertainty, and we can trust it
only at the order of magnitude level.  We have explored the sensitivity of this
qualitative result to a number of the assumptions we have made, and have found
that changing the details of the model does not significantly alter our
conclusions.  Furthermore, including some $SU(3)$ breaking in the matrix
elements cancels to some extent in $x/y$ and does not induce dramatic changes.

The significance of our result is clear only in the context of the experimental
situation.  The current bounds on $x$ and $y$ are at the level of a few
percent, and the central question is whether their actual observation at or
just below this level could be interpreted as a clear signal of physics beyond
the Standard Model.  We would argue that our analysis has taught us that,
without further refinement, the answer is no.  We have identified a real effect
that could plausibly give $x$ and $y$ at the percent level, albeit with very
large uncertainties.

In general, an observation of $x \gg y$ would be an indication for new physics,
but this could only be established if $y$ were very small, at the $10^{-3}$
level.  Such a situation could arise if new physics enhanced $x$ but not $y$. 
Yet since one cannot exclude the possibility of cancellation between different
SM contributions to $y$, even this outcome would not admit an unambiguous
interpretation.  

However, if $x$ were indeed enhanced by new physics, such new physics may also
introduce a sizable new $CP$ violating phase which may be observable.  Thus, we
would argue that in $D^0-\D0bar$ mixing, the only single measurement that could
establish by itself the presence of new physics would be the observation of
$CP$ violation, which is very small in the Standard Model independent of
hadronic effects.

\acknowledgments

It is a pleasure to thank Mark Wise, as usual, for helpful conversations.  
We are grateful to the Aspen Center for Physics for its hospitality while
portions of this work were completed.
A.F.~was supported in part by the U.S.~National Science Foundation under Grant
PHY--9970781. 
Y.G.~was supported in part by the Department of Energy, contract
DE-AC03-76SF00515 and by the Department of Energy under grant
no.~DE-FG03-92ER40689.
Z.L.~was supported in part by the Director, Office of Science, Office of
High Energy and Nuclear Physics, Division of High Energy Physics, of the U.S.\
Department of Energy under Contract DE-AC03-76SF00098 and by a DOE Outstanding
Junior Investigator award.
The work of Y.G.\ and Z.L.\ was also supported in part by the United
States--Israel Binational Science Foundation (BSF) through Grant No.\ 2000133.
Y.N.~is supported by the Israel Science Foundation founded by the Israel
Academy of Sciences and Humanities, by EEC RTN contract HPRN-CT-00292-2002, by
a Grant from the G.I.F., the German-Israeli Foundation for Scientific Research
and Development, by the Minerva Foundation (M\"unchen), and by a grant from the
United States-Israel Binational Science Foundation (BSF), Jerusalem, Israel. 
A.P.~was supported in part by the U.S.\ National Science Foundation under
Grant PHY--0244853, and by the U.S.\ Department of Energy under Contract
DE-FG02-96ER41005.

\end{document}